\documentclass[12pt,preprint]{aastex}

\usepackage{epsfig}
\shorttitle{Electron temperatures from \ion{He}{1} lines}
\shortauthors{Zhang et al.}

\begin{document}

\title{Helium recombination spectra as temperature diagnostics for planetary
nebulae}

\author{Y. Zhang \altaffilmark{1}} 
\affil{Department of Astronomy, Peking University, Beijing 100871,
        P. R. China}
\email{zhangy@bac.pku.edu.cn}

\author{X.-W. Liu}
\affil{Department of Astronomy, Peking University, Beijing 100871,
        P. R. China}
\author{Y. Liu}
\affil{Department of Physics, Guangzhou University, Guangzhou 510405, P. R. China}
\author{R. H. Rubin \altaffilmark{2}}
\affil{NASA Ames Research Center, Moffett Field, CA 94035-1000, USA}

\altaffiltext{1}{Present address: Space Telescope Science Institute,
3700 San Martin Drive, Baltimore, MD 21218}
\altaffiltext{2}{Orion Enterprises, M.S. 245-6, Moffett Field, CA 94035-1000, USA}

\begin{abstract} Electron temperatures derived from the \ion{He}{1}
recombination line ratios, designated $T_{\rm e}$(\ion{He}{1}), are presented
for 48 planetary nebulae (PNe). We study the effect that temperature
fluctuations inside nebulae have on the $T_{\rm e}$(\ion{He}{1}) value. We show
that a comparison between $T_{\rm e}$(\ion{He}{1}) and the electron temperature
derived from the Balmer jump of the \ion{H}{1} recombination spectrum,
designated $T_{\rm e}$(\ion{H}{1}), provides an opportunity to discriminate
between the paradigms of a chemically homogeneous plasma with temperature and
density variations, and a two-abundance nebular model with hydrogen-deficient
material embedded in diffuse gas of a ``normal'' chemical composition (i.e.
$\sim$ solar), as the possible causes of the dichotomy between the abundances
that are deduced from collisionally excited lines to those deduced from
recombination lines. We find that $T_{\rm e}$(\ion{He}{1}) values are
significantly lower than $T_{\rm e}$(\ion{H}{1}) values, with an average
difference of $\langle T_{\rm e}$(\ion{H}{1})-$T_{\rm
e}$(\ion{He}{1})$\rangle=4000$\,K. The result is consistent with the
expectation of the two-abundance nebular model but is opposite to the
prediction of the scenarios of temperature fluctuations and/or density
inhomogeneities. From the observed difference between $T_{\rm e}$(\ion{He}{1})
and  $T_{\rm e}$(\ion{H}{1}), we estimate that the filling factor of
hydrogen-deficient components has a typical value of $10^{-4}$.  In spite of
its small mass, the existence of hydrogen-deficient inclusions may potentially
have a profound effect in enhancing the intensities of \ion{He}{1}
recombination lines and thereby lead to apparently overestimated helium
abundances for PNe.

\end{abstract}


\section{Introduction}

Two long-standing problems in nebular astrophysics are a) heavy element
abundances relative to hydrogen obtained from collisionally excited lines
(CELs) are generally lower than those determined from optical recombination
lines (ORLs); and b) electron temperatures deduced from the collisionally
excited [\ion{O}{3}] nebular-to-auroral forbidden line ratio -- hereafter
$T_{\rm e}$([\ion{O}{3}]) -- are systematically higher than those determined
from the Balmer jump (BJ) of \ion{H}{1} recombination spectrum -- hereafter $T_{\rm
e}$(\ion{H}{1}) -- (see \citealt{liu209} for a recent review).

Several solutions have been proposed for the discrepancies.  \citet{peim67}
showed that temperature fluctuations within nebulae can lead to a $T_{\rm
e}$([\ion{O}{3}]) which overestimates the average emission temperature of 
the \ion{H}{1} recombination spectrum and the [\ion{O}{3}] nebular lines. As a
consequence, the O$^{2+}$/H$^+$ ionic abundance ratio derived from the
intensity ratio of the [\ion{O}{3}] nebular lines to H$\beta$, assuming $T_{\rm
e}$([\ion{O}{3}]) as the electron temperature, may be grossly underestimated.
The presence of temperature fluctuations will also lead to a $T_{\rm
e}$([\ion{O}{3}]) that is systematically higher than $T_{\rm e}$(\ion{H}{1}),
the temperature derived from the \ion{H}{1} recombination spectrum.  Evidence
in favor of this was first presented by \cite{peim71} who measured values of
$T_{\rm e}$(\ion{H}{1}) in several H~{\sc ii} regions and planetary nebulae
(PNe) from the observed magnitude of the H~{\sc i} Balmer jump and found that they
are systematically lower than the corresponding values of $T_{\rm
e}$([\ion{O}{3}]) of the same nebulae.  A parameter $t^2$ was introduced by
\citet{peim67} to characterize the magnitude of temperature fluctuations in a
given nebula.  The scenario has however been found to be unable to account for
the low abundances derived from temperature-insensitive IR CELs.   High
quality measurements of IR fine-structure lines with the {\it Infrared Space
Observatory}\ ({\it ISO}) for a number of PNe have shown that these lines of
very low excitation energies generally yield abundances similar to the values
inferred from optical/UV CELs. Again they are lower with respect to ORL
abundances \citep[e.g.][]{liubarlow2000,liuyi04b,tsamis04}. Moreover, recent
observational studies of several samples of PNe show that typical values of
$t^2$ derived by comparing $T_{\rm e}$([\ion{O}{3}]) and $T_{\rm
e}$(\ion{H}{1}) are so large that they are well beyond the predictions of any
photoionization models \citep{liu93,zhang2004} unless extra energy
input (such as from magnetic reconnection or from shocks) is considered.

A more plausible solution for the discrepancies of temperature and abundance
determinations using CELs and ORLs is the two-component nebular model with
hydrogen-deficient inclusions embedded in the diffuse nebula first proposed by
\cite{liubarlow2000}. Unlike the purely phenomenological temperature
fluctuations scenario, the two-component nebular model provides a physical
explanation for the temperature and abundance determination discrepancies.  The
model assumes that there is a small amount of H-deficient material embedded in
the diffuse nebula of ``normal'' composition. Due to the much enhanced cooling
rates by IR fine-structure lines of heavy element ions as a result of the high
metallicity, hydrogen-deficient inclusions have so low an electron temperature
that they emit copiously in ORLs but essentially not at all in CELs.  In this
scenario, the high heavy elemental abundances yielded by ORLs simply reflect
the enhanced metallicity in those H-deficient inclusions, rather than
representing the overall chemical composition of the whole nebula.  Detailed
two-abundance photoionization models incorporating H-deficient inclusions have
been constructed by \citet{pequignot}.  The models provide a satisfying
explanation for the apparent large discrepancies between the results of
temperature and abundance determinations using CELs and ORLs, for the PNe
\object{NGC\,6153}, \object{M\,1-42} and \object{M\,2-36}.  
Contrary to the temperature fluctuations scenario, the two-component nebular
model does not need any extra energy input to reproduce the large temperature 
inhomogeneities within PNe.  On the other hand,
the presence of such H-deficient inclusions is not predicted by any of the
current theories of the nucleosynthesis and evolution of low- and
intermediate-mass stars and their nature and origin remain unclear. One
possibility is that they are evaporating icy planetesimals (Liu 2003).

More plasma diagnostic methods are helpful to probe nebular physical
structure and to advance our understanding of the aforementioned fundamental
problems. Earlier studies \citep[e.g.][]{peimbert2000,peimbert2002} show that
temperature fluctuations can affect the intensities of \ion{He}{1} lines.
\citet{peimbert1995} presented a method to determine the average emission
temperature of \ion{He}{1} lines, -- hereafter $T_{\rm e}$(\ion{He}{1}) -- by
adjusting the optical depth, the $t^2$ parameter and electron temperature to
reconcile the He$^+$/H$^+$ ratios yielded by individual  \ion{He}{1} lines.
The method however requires accurate atomic data in
addition to high quality line intensity measurements.

Increasingly reliable atomic data are becoming available. A calculation
for the effective recombination coefficients of a selection of \ion{He}{1}
lines was given by \citet{smits1996}.  \citet{sawey} studied the contributions
of collisional excitation from the He$^0$ 2s$^3$S and 2s$^1$S meta-stable
levels by electron impacts to the observed fluxes of \ion{He}{1} lines.
Combining the recombination data of \citet{smits1996} and the collision
strengths of \citet{sawey}, \citet{benjamin} presented improved \ion{He}{1}
line emission coefficients and fit the results with analytical formulae.  More
recently, \citet{benjamin2002} studied the effects of the optical depth of the
2s$^3$S meta-stable level on \ion{He}{1} line intensities. The availability of
these improved atomic data has made it possible to obtain secure measurements
of $T_{\rm e}$(\ion{He}{1}), given high quality spectroscopic data.

The main goal of the current paper is to demonstrate that \ion{He}{1} emission
lines provide a valuable probe of nebular thermal structure and a method to
discriminate different scenarios proposed as the causes of the CEL versus ORL
temperature and abundance determination discrepancies.  In Section 2, we
present electron temperatures derived from \ion{He}{1} lines and confront the
results with the predictions of the scenarios of temperature fluctuations,
density inhomogeneities and of the two-abundance nebular model in Section 3. We
summarize our results in Section 4.

\section{Electron temperatures from \ion{He}{1} lines}

\subsection{Method and atomic data}

\citet{benjamin} provide analytic formulae for the emissivities of \ion{He}{1}
lines as a function of electron temperature for $N_{\rm e} = 10^2, 10^4$ and
$10^6$\,cm$^{-3}$, including contributions from recombinations of He$^+$ with
electrons \citep{smits1996} and from the effects of collisional excitation from
the 2s$^3$S and 2s$^1$S meta-stable levels \citep{sawey}.  Using their
formulae, the intensity ratio of two \ion{He}{1} lines is given by
\begin{equation}
  \frac{I_1}{I_2}=\frac{a_1}{a_2}T_{\rm e4}^{b_1-b_2}
  \exp(\frac{c_1-c_2}{T_{\rm e4}}),
\end{equation}
where $T_{\rm e4}=T_{\rm e}/10^4$\,K.  Values of the fitting parameters $a_i$,
$b_i$ and $c_i$ are given by \citet{benjamin} for the temperature range
5\,000--20\,000\,K, within which the emissivity fits have a maximum error of
less than 5\%.  For lower temperatures, we have obtained similar fits using the
numerical values given by \citet{smits1996} and \citet{sawey} and present the
results in Appendix~A. For singlet lines, Case B recombination was assumed.

\begin{figure*}
\epsfig{file=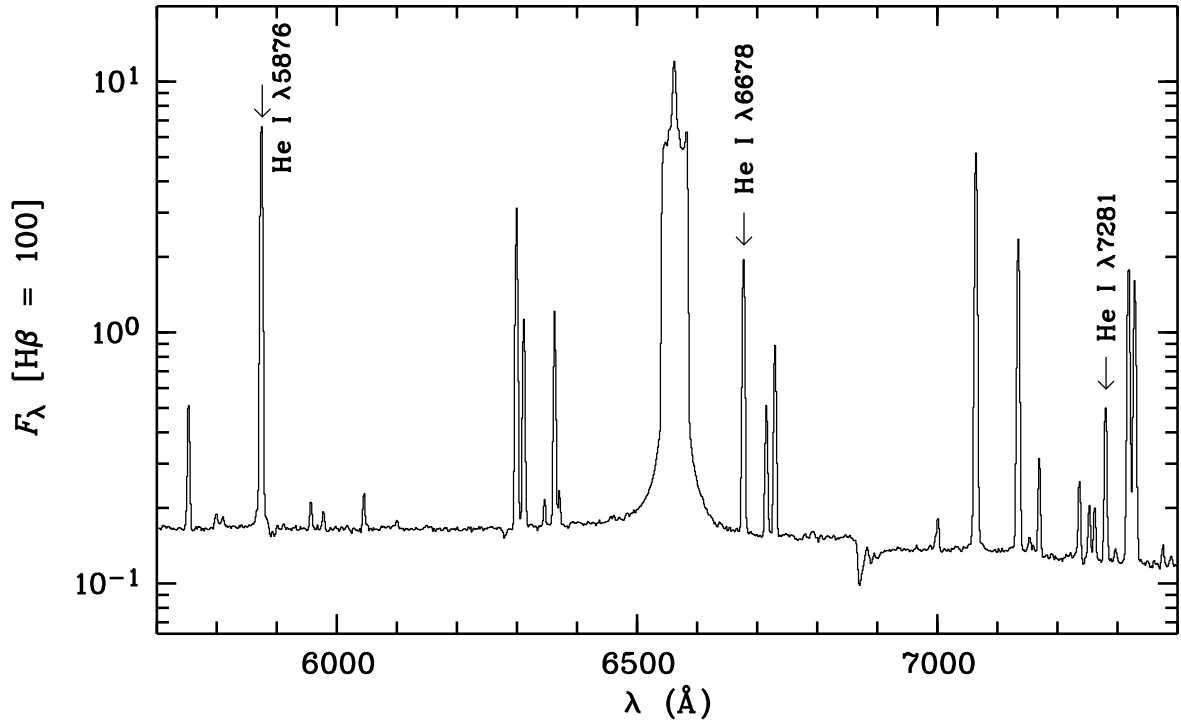,
height=10.5cm, bbllx=73, bblly=268, bburx=523, bbury=553, clip=, angle=0}
\caption{Spectrum of NGC\,7009 showing the \ion{He}{1} $\lambda\lambda$5876, 
6678 and 7281 lines used to determine average emission temperatures of the 
\ion{He}{1} recombination spectrum. The intensity was scaled such that 
$F({\rm H}\beta)=100$. Note that H$\alpha$ was heavily saturated in this 
spectrum.}
\label{spec}
\end{figure*}

By measuring the intensity ratio of two \ion{He}{1} recombination lines, one
can in principle determine the electron temperature from Eq.~(1).  This is
however not an easy task given the relatively weak dependence of the line ratio
on temperature. Apart from recombination and collision excitation, other minor
effects that can potentially affect the observed line intensities may also need
to be considered, such as self-absorption from the meta-stable 2s\,$^3$S level
which can be significantly populated under typical nebular conditions.  In
order to achieve the best measurements of the average \ion{He}{1} line emission
temperature, one needs to take into account a variety of considerations,
including the strengths and ability to measure the lines involved in the ratio,
the accuracy and reliability of the atomic data for the lines, the sensitivity
of the ratio to $T_{\rm e}$, and the magnitudes of contaminations by other
unwanted side effects such as collisional excitation and self-absorption.
After a thorough investigation, we decided that the best line ratio suitable
for temperature determinations is probably the $\lambda7281$/$\lambda6678$, for
the following reasons: 1) The $\lambda\lambda7281,6678$ lines are amongst the
strongest \ion{He}{1} lines observable in the optical, after the $\lambda$5876
and $\lambda$4471 lines, and both lines fall in a spectral region clear of
serious telluric absorption/emission and blending by any other known strong
spectral features (see Fig.~\ref{spec} for a typical nebular spectrum showing
the relevant spectral region); 2) The atomic data associated with the
$\lambda\lambda7281,6678$ lines are amongst the best determined; 3) The
\ion{He}{1} $\lambda7281$/$\lambda6678$ ratio is more sensitive to temperature
but less sensitive to density compared to other ratios of strong \ion{He}{1}
lines; 4) Both the $\lambda6678$ and $\lambda7281$ lines are from singlet
states and thus are essentially unaffected by the optical depth effects of the
2s\,$^3$S level; 5) Given the small wavelength span between the
$\lambda\lambda7281,6678$ lines, measurements of their intensity ratio are less
sensitive to any uncertainties in reddening corrections and flux calibration
\citep[c.f. also][]{liuyi04b}.

\subsection{Results}

In this work, a total of 48 PNe are analyzed. Amongst them, 23 were observed
with the 4.2\,m William Herschel Telescope at La Palma and the ESO 1.52\,m
telescope, as described in \citet{zhang2004}.  For the remaining sources,
spectral data are from \citet{liuyi04a}, \citet{tsamis}, \citet{wesson04},
\citet{ruiz2003} and \citet{peimbert2004}.  The corresponding references are
listed in the last column of Table~\ref{result}.

\begin{figure*}
\epsfig{file=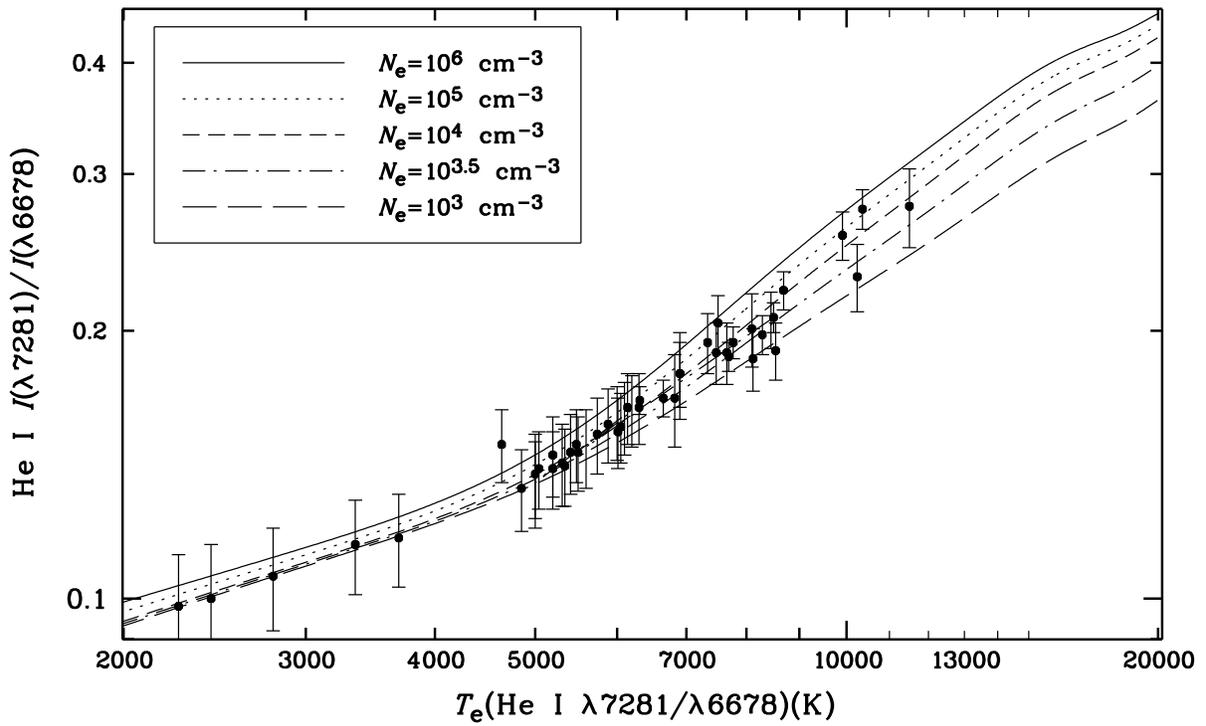,
height=10cm, bbllx=61, bblly=278, bburx=533, bbury=565, clip=, angle=0}
\caption{The \ion{He}{1} $\lambda7281/\lambda6678$ intensity ratio as a
function of $T_{\rm e}$ for electron densities from 
$N_{\rm e}=10^3$ to 10$^6$\,cm$^{-3}$. The observed ratios along with their
uncertainties for our sample PNe are overplotted.}
\label{ra_te}
\end{figure*}

In Fig.~\ref{ra_te}, we plot the \ion{He}{1} $\lambda7281$/$\lambda6678$ ratio
versus $T_{\rm e}$(\ion{He}{1} $\lambda7281$/$\lambda6678$) for different electron densities. The observed \ion{He}{1}
$\lambda7281$/$\lambda6678$ ratios along with measurement uncertainties are
also overplotted. The electron densities of individual nebulae are taken from
\citet{zhang2004}, \citet{liuyi04a}, \citet{tsamis}, \citet{wesson04},
\citet{ruiz2003} and \citet{peimbert2004}.  The observed line fluxes have been
corrected for dust extinction using reddening constants taken from the same
references. The resultant electron temperatures derived from the
$\lambda7281$/$\lambda6678$ ratio are presented in Table~\ref{result}.  The
associated uncertainties were estimated based on the line ratio measurement
errors.  Inspection of Fig.~\ref{ra_te} shows that most PNe in our sample have
$T_{\rm e}$(\ion{He}{1}) less than $10\,000$\,K. For this low temperature
regime, the $\lambda7281$/$\lambda6678$ ratio is quite insensitive to electron
density.  Thus our temperature determinations should be hardly affected by any
uncertainties in density measurements.

\begin{figure*}
\epsfig{file=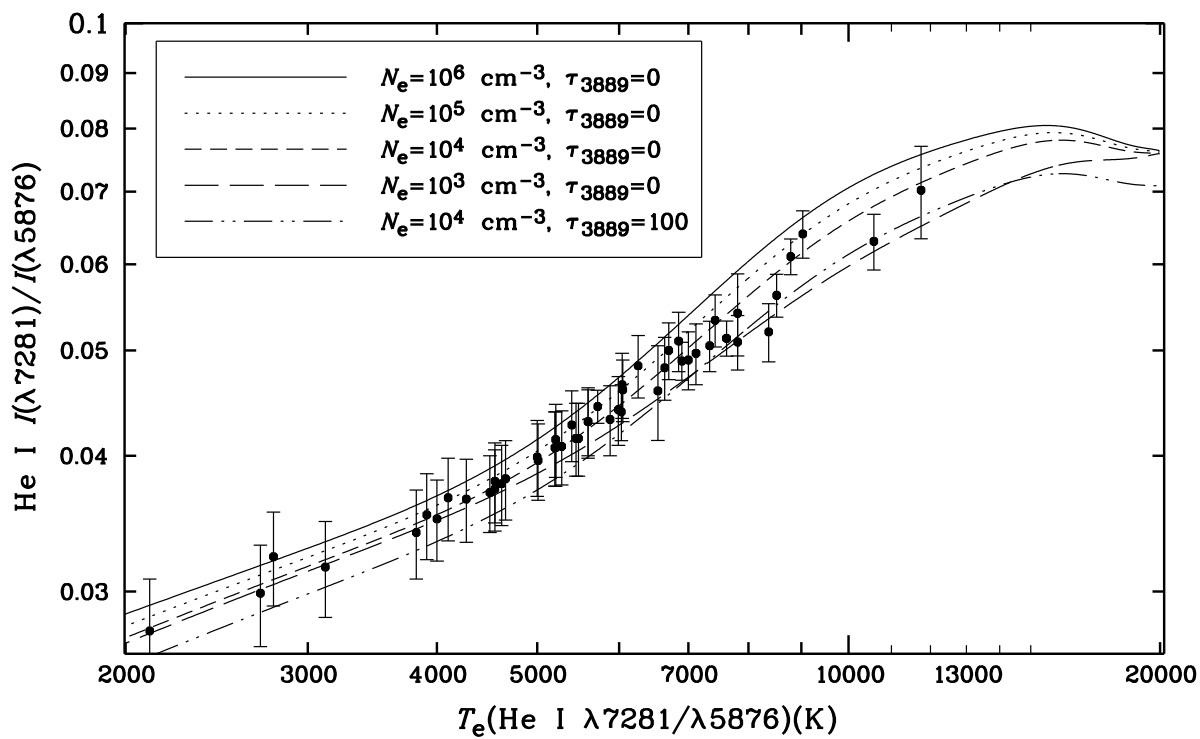,
height=10cm, bbllx=61, bblly=278, bburx=533, bbury=565, clip=, angle=0}
\caption{The \ion{He}{1} $\lambda7281/\lambda5876$ intensity ratio is plotted 
against $T_{\rm e}$ for densities from $N_{\rm e}=10^3$ to 10$^6$\,cm$^{-3}$. 
All curves are for zero $\lambda$3889 line optical depth except one
for which $\tau_{3889}=100$ and $N_{\rm e}=10^4$\,cm$^{-3}$ (see text for 
more details).}
\label{ra_te2}
\end{figure*}

For comparison, we have also determined temperatures using the \ion{He}{1}
$\lambda7281$/$\lambda5876$ ratio.  For this ratio, additional uncertainties
may arise due to the optical depth effects of the 2s\,$^3$S level on the
intensity of the $\lambda5876$ line.  Based on the line emissivities given by
\citet{benjamin2002}, we have also plotted in Fig.~\ref{ra_te2} one curve
showing the variations of the $\lambda7281$/$\lambda5876$ ratio as a function
of $T_{\rm e}$ for the case of $N_{\rm e}=10^4$\,cm$^{-3}$ and $\tau_{3889} =
100$, the optical depth at the centre of the \ion{He}{1} 2s\,$^3$S --
3p\,$^3$P$^{\rm o}$ $\lambda$3889 line.  Fig.~\ref{ra_te2} shows that
uncertainties in $\tau_{3889}$ may in principle lead to an error of
approximately 1000\,K in temperature determinations and that the uncertainty
increases with increasing temperature.  Values of $T_{\rm e}$(\ion{He}{1})
derived from the $\lambda7281$/$\lambda5876$ ratio are also tabulated in
Table.~\ref{result}. Possible errors caused by uncertainties in $\tau_{3889}$
have not been included in the error estimates.

\begin{figure*}
\epsfig{file=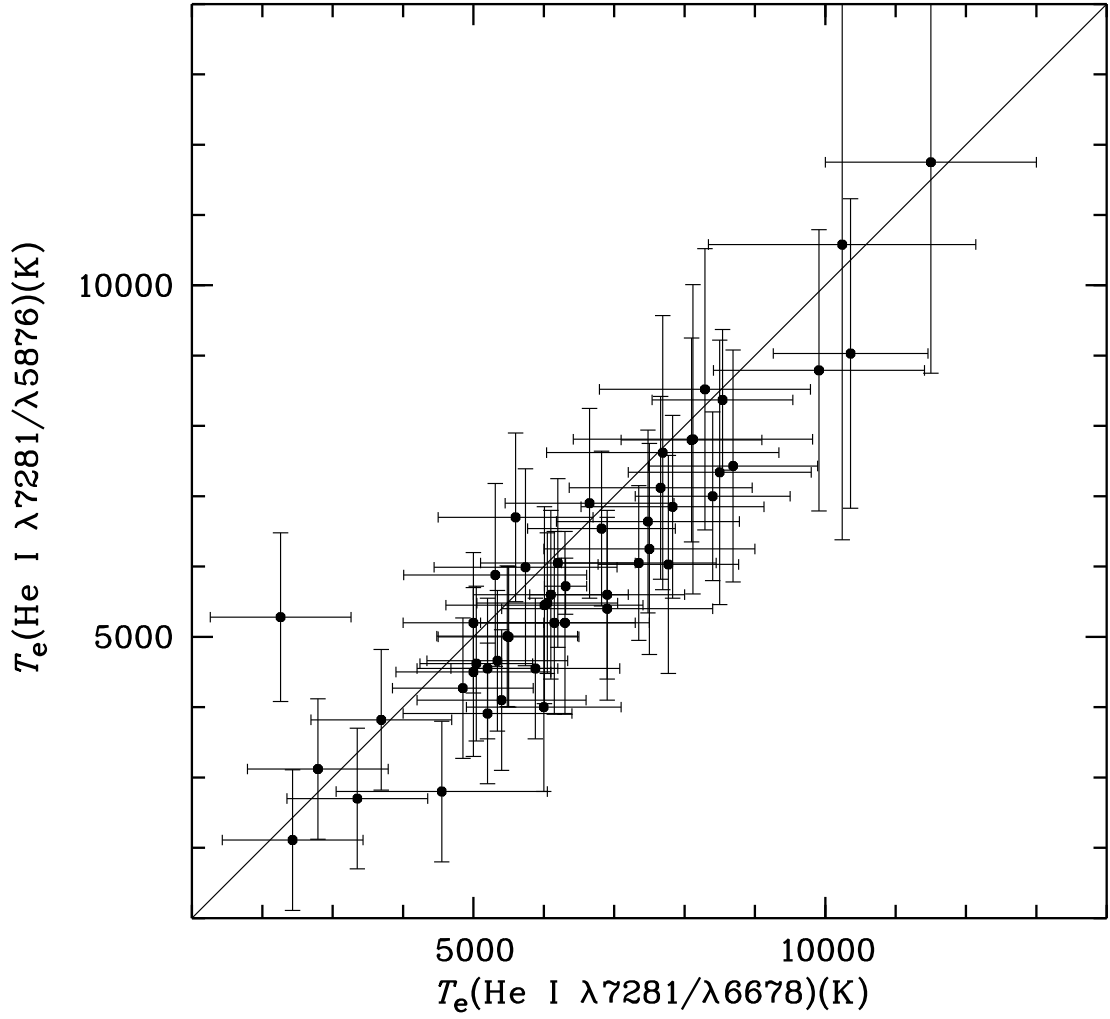,
height=14cm, bbllx=61, bblly=273, bburx=549, bbury=697, clip=, angle=0}
\caption{Comparison of electron temperatures derived from
the \ion{He}{1} $\lambda7281/\lambda6678$ ratio with those derived
from the \ion{He}{1} $\lambda7281/\lambda5876$ ratio. The diagonal line
is a $y=x$ plot.}
\label{hehe}
\end{figure*}

In Fig.~\ref{hehe}, we compare electron temperatures derived from the
$\lambda7281/\lambda6678$ ratio and from the $\lambda7281/\lambda5876$ ratio.
Inspection of the figure shows that temperatures derived from the two
ratios are generally in good agreement. There is however some evidence that
the $\lambda7281/\lambda5876$ ratio tends to yield lower temperatures.  Two
effects may be responsible for the small offsets. In our calculations, the
possible effects of self-absorption on the intensity of the $\lambda5876$ line
were not considered.  If $\tau_{3889}>0$, then our calculations may have
systematically underestimated the $\lambda7281/\lambda5876$ temperatures (c.f.
Fig.~\ref{ra_te2}).  Alternatively, if there is some small departure from pure
Case B to Case A recombination for the \ion{He}{1} singlet lines, then our
calculations may have underestimated temperatures derived from both ratios.
But this effect is expected to be small for temperatures derived from the
$\lambda7281/\lambda6678$ ratio as both lines involved in the ratio are from
singlet spectral terms, and as a result the effect may have largely canceled
out.  It is difficult to discriminate between these two possibilities. In any
case, the relatively small offset between the two temperatures as shown in
Fig.~\ref{hehe} suggests that both effects are not large and should not have
affected our temperature determinations by a significant amount.

Table~\ref{result} also lists values of $T_{\rm e}$(\ion{H}{1}), the
temperature derived from the ratio of the hydrogen Balmer discontinuity at
3646\,{\AA} to H11 $\lambda3770$, taken from \citet{zhang2004}. A comparison
between $T_{\rm e}$(\ion{He}{1}) and $T_{\rm e}$(\ion{H}{1}) is given in
Fig.~\ref{heihi}.  It shows that, with the only exception of \object{NGC\,40},
$T_{\rm e}$(\ion{He}{1}) is lower than $T_{\rm e}$(\ion{H}{1}). The average
difference between the two temperatures for the whole sample is $\langle T_{\rm
e}$(\ion{H}{1})$-$$T_{\rm e}$(\ion{He}{1})$\rangle$=$4000$\,K.

\begin{figure*}
\epsfig{file=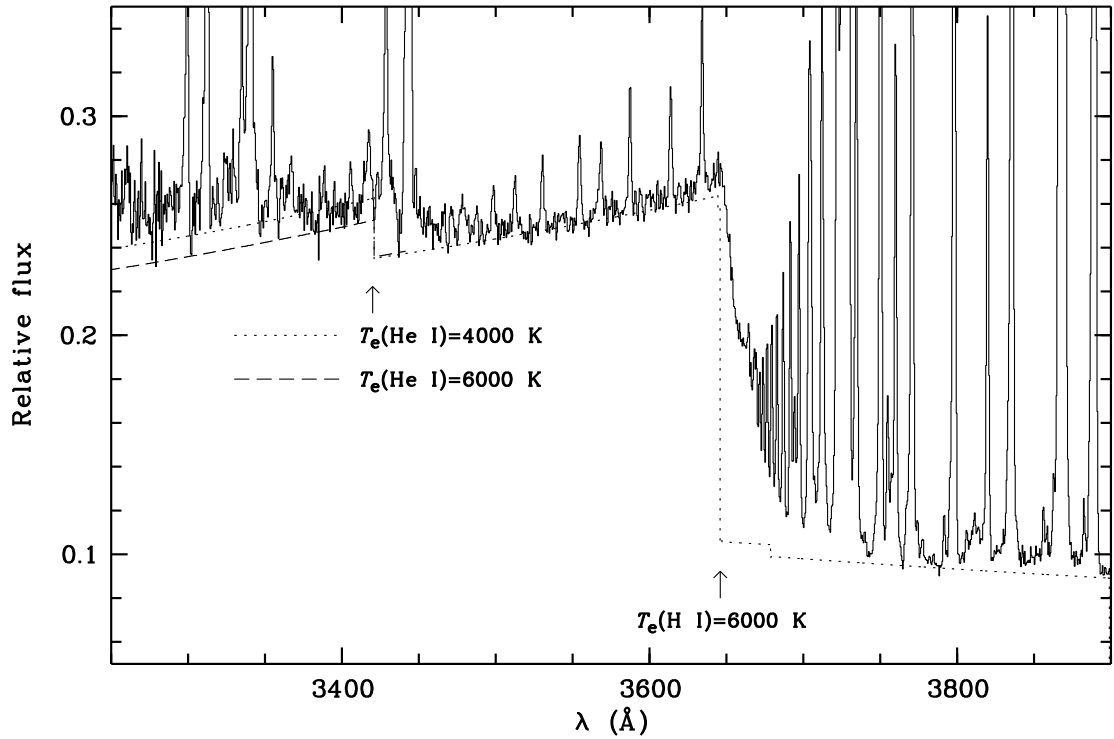,
height=10cm, bbllx=60, bblly=276, bburx=548, bbury=600, clip=, angle=0}
\caption{Spectrum of NGC\,6153 from 3250--3900\,{\AA}, showing the \ion{He}{1}
discontinuity at 3421\,{\AA} and the \ion{H}{1} discontinuity at 3646\,{\AA}.
The dotted line is a theoretical nebular continuum spectrum calculated assuming
a \ion{He}{1} temperature of 4000\,K and a \ion{H}{1} temperature of 6000\,K.
 For comparison, a theoretical spectrum assuming a \ion{He}{1} temperature
of 6000\,K is also given (dashed line), which clearly yields a poorer fit to 
the observed magnitude of the \ion{He}{1} $\lambda$3421 discontinuity than the
dotted line.}
\label{jum}
\end{figure*}

A measure of the electron temperature characterizing the \ion{He}{1} 
zones can also be obtained by observing the \ion{He}{1}
discontinuity at 3421\,{\AA}, produced by He$^{+}$ recombinations to the
\ion{He}{1} 2p\,$^3$P$^{\rm o}$ level \citep{liu93}. With $T_{\rm
e}$(\ion{H}{1}) and $T_{\rm e}$(\ion{He}{1}) as free parameters, the \ion{H}{1}
and \ion{He}{1} recombination continuum spectrum is calculated to fit the
observed continuum spectrum of \object{NGC\,6153}. The best fit yields
$T_{\rm e}$(\ion{He}{1}) = $4000\pm1500$\,K (Fig.~\ref{jum}), consistent within
the uncertainties with the value $3350\pm1000$\,K derived from the \ion{He}{1}
$\lambda7281$/$\lambda6678$ line ratio. The \ion{He}{1} $\lambda$3421
discontinuity is much weaker than the \ion{H}{1} Balmer discontinuity and falls
in a spectral region of even shorter wavelengths crowded by strong emission
lines such as the \ion{O}{3} Bowen fluorescence lines
$\lambda\lambda$3428,3444, and is therefore much more difficult to measure.
Nevertheless, high quality spectroscopic observations of this discontinuity
should be invaluable. 

\section{Discussion}

\subsection{Temperature fluctuations and density inhomogeneities}

The discrepancy between electron temperatures derived from the [\ion{O}{3}]
forbidden line ratio and those deduced from the hydrogen Balmer jump was first
discovered by \citet{peim71} and was ascribed to temperature fluctuations
within nebulae. For a nebula with small magnitude temperature
fluctuations, for a given ionic species of number density $N_i$, the nebular
thermal structure can be characterized by an average temperature $T_0$ and a
mean square temperature fluctuation parameter $t^2$ defined as \citep{peim67},
\begin{equation}
T_0(N_i)=\frac{\int T_{\rm e}N_{\rm e}N_idV}{\int N_{\rm e}N_idV}
\end{equation}
and
\begin{equation}
t^2(N_i)=\frac{\int (T_{\rm e}-T_0)^2N_{\rm e}N_idV}{T_0^2\int N_{\rm e}N_idV}.
\end{equation}

For $T_{\rm e}$(\ion{H}{1}) derived from the H~{\sc i} recombination spectrum
Balmer discontinuity, it can be shown that \citep{peim67},
\begin{equation}
T_{\rm e}({\rm H~{\sc I}})=T_0(1-1.67t^2).
\end{equation}
The deduction implicitly assumes that $t^2\ll 1$.

Adopting the analytic expression for the fit of line emissivity as described in
Section~2.1, we can write the flux of a \ion{He}{1} recombination line as,
\begin{equation}
I({\rm He~{\sc I}},\lambda_i)=\int N_{\rm e}N({\rm He^+})a_iT_{{\rm e}4}^{b_i}\exp(\frac{c_i}{T_{\rm e4}})dV.
\end{equation}

Then if $t^2\ll1$, one can rewrite the expression as a Taylor series expanded
around the average temperature $T_0$, keeping only terms to the second order
\citep{peim67},

\begin{eqnarray}
\nonumber
I({\rm He\,{\sc I}},\lambda_i)\approx a_iT_{04}^{b_i}\exp\left(\frac{c_i}{T_{04}}\right)\int N_{\rm e}N({\rm He^+})\hspace{3.1in} \\
\nonumber
\times\left\{1+\left(b_i-\frac{c_i}{T_{04}}\right)\left(\frac{T_{\rm e}-T_0}{T_0}\right)
+\frac{1}{2}\left[b_i(b_i-1)-\frac{2c_i(b_i-1)}{T_{04}}
+\frac{c_i^2}{T_{04}^2}\right]\left(\frac{T_{\rm e}-T_0}{T_0}\right)^2\right\}dV,\\
\end{eqnarray}
where $T_{04}=T_0/10^4$\,K. Then from Eqs.~(2), (3) and (6), we have, 
\begin{eqnarray} 
\nonumber
\frac{I({\rm He~{\sc I}},\lambda_1)}{I({\rm He~{\sc I}},\lambda_2)}&\equiv&\frac{a_1}{a_2}T_{\rm e4}({\rm He~{\sc I}},\lambda_1/\lambda_2)^{b_1-b_2}\exp\left[\frac{c_1-c_2}{T_{\rm e4}({\rm He~{\sc I}},\lambda_1/\lambda_2)}\right]\\
 &\approx & \frac{a_1}{a_2}T_{04}^{b_1-b_2}\exp\left(\frac{c_1-c_2}{T_{04}}\right)[1+A(T_0)t^2],
\end{eqnarray}
where
\begin{equation}
A(T_0)=\frac{1}{2}\left\{b_1(b_1-1)-b_2(b_2-1)-2[c_1(b_1-1)-c_2(b_2-1)]\frac{1}{T_{04}}+\frac{c_1^2-c_2^2}{T_{04}^2}\right\}.
\end{equation}
Introducing a parameter $B$ and when $|Bt^2|\ll 1$, we can obtain from
Eqs.~(7)
\begin{eqnarray}
\left\{
\begin{array}{l}
T_{\rm e4}({\rm He~{\sc I}},\lambda_1/\lambda_2)^{b_1-b_2}\exp\left[\frac{c_1-c_2}{T_{\rm e4}({\rm He~{\sc I}},\lambda_1/\lambda_2)}\right]
 \approx \left[T_{04}(1-Bt^2)\right]^{b_1-b_2}\exp\left[\frac{c_1-c_2}{T_{04}(1-B t^2)}\right]\\
B=T_{04}A(T_0)/[c_1-c_2-(b_1-b_2)T_{04}].
\end{array}
\right.
\end{eqnarray}
Thus if $t^2\ll1$ and $|Bt^2|\ll 1$, we have
\begin{equation}
T_{\rm e}({\rm He~{\sc I}},\lambda_1/\lambda_2)\approx T_0\left[1-\frac{T_0A(T_0)}{10^4(c_1-c_2)-(b_1-b_2)T_0}t^2\right].
\end{equation}

In Eq.\,(10) the coefficient in front of $t^2$ is a slowly varying function of
$T_0$.  For $T_{\rm e}$(\ion{He}{1}; $\lambda_1/\lambda_2$) derived from the
$\lambda7281$/$\lambda6678$ ratio, according to the \ion{He}{1} line emissivity
fit parameters $b_1$, $b_2$, $c_1$ and $c_2$ given by \citet{benjamin} for
$T_{\rm e}$ between 5000 and 20\,000~K and $N_{\rm e}=10^4$~cm$^{-3}$, this
coefficient varies between 0.97 and 1.25.  For $T_{\rm e} < 5000$~K and $N_{\rm
e}=10^4$~cm$^{-3}$, using the fit parameters given in Appendix~A
(Table~\ref{fit}), we find that this coefficient varies between 1.03 and 1.22.
Therefore, as an approximation, we adopt a value of 1.07 for this coefficient,
as calculated for $T_{\rm e}=10000$~K and $N_{\rm e}=10^4$~cm$^{-3}$, and have
\begin{equation} T_{\rm e}({\rm He~{\sc I}},\lambda7281/\lambda6678)\approx
T_0(1- 1.07t^2).  \end{equation}

Under the scenario of small temperature fluctuations ($t^2\ll 1$), we can
reasonably assume that $T_{\rm 0}$(\ion{He}{1}) $\approx T_{\rm 0}$(\ion{H}{1})
$ = T_{\rm 0}$ (c.f. Section~3.2 for further discussion on the legitimacy of
this assumption) and $t^2$(\ion{He}{1}) $\approx t^2$(\ion{H}{1}) $ = t^2$.
Consequently, a comparison of Eqs.~(4) and (11) shows that $T_{\rm
e}$(\ion{He}{1}, $\lambda7281/\lambda6678$) $\ga$ $T_{\rm e}$(\ion{H}{1}),
which is exactly opposite to what is observed (c.f.  Section~2.2,
Table~\ref{result}).

Using Eqs.~(4) and (10), we plot in Fig.~\ref{heihi} $T_{\rm e}({\rm
H~{\sc I}})$ as a function of $T_{\rm e}$(\ion{He}{1},
$\lambda7281/\lambda6678$), for the case of $t^2=0.00$, 0.04, 0.10 and 0.16,
assuming a constant density of 10$^4$\,cm$^{-3}$. Obviously, the differences
between $T_{\rm e}({\rm H~{\sc I}})$ and $T_{\rm e}({\rm He~{\sc I}})$ are in
the opposite direction to what is predicted by temperature fluctuations.  
It follows that the \ion{He}{1} lines may arise from regions characterized by
significantly lower temperature than that of the \ion{H}{1} emission regions,
i.e., $T_{\rm 0}$(\ion{He}{1})$<T_{\rm 0}$(\ion{H}{1}). This, however, will
invalidate the condition $t^2\ll 1$, and thus is beyond the scope of the
description of the temperature fluctuations as originally envisioned by
\citet{peim67,peim71}. A detailed discussion will be given below in
Section~3.2.

Another theory that has been proposed to explain the ORL/CEL abundance
determination and the BJ/CEL temperature determination discrepancies is density
variations in a chemically homogeneous nebula \citep{rubin1989,viegas}.  In
this scenario, because the [\ion{O}{3}] $\lambda$4363 auroral line has a much
higher critical density than the $\lambda\lambda$4959,5007 nebular lines and is
therefore less affected by collisional de-excitation in high density regions,
the presence of such high density regions will lead to an overestimated $T_{\rm
e}$([\ion{O}{3}]) and consequently underestimated CEL abundances, analogous to
the effects of temperature fluctuations. The presence of moderate density
inhomogeneities in PNe is confirmed by recent observations
\citep[e.g.][]{zhang2004}. However, given that \ion{H}{1} and \ion{He}{1}
recombination lines are permitted transitions and therefore hardly
affected by collisional de-excitation, such a scenario predicts $T_{\rm
e}$(\ion{He}{1})$\sim T_{\rm e}$(\ion{H}{1}) (as long as \ion{He}{1} zone $\sim$
\ion{H}{1} zone; see discussion below in section~3.2), also inconsistent with observations. We thus conclude that for
a chemically homogeneous nebula, the possible presence of large density
inhomogeneities also fails to explain the observed differences between $T_{\rm
e}$(\ion{H}{1}) and $T_{\rm e}$(\ion{He}{1}).

\subsection{Two-abundance nebular models}

\begin{figure*}
\epsfig{file=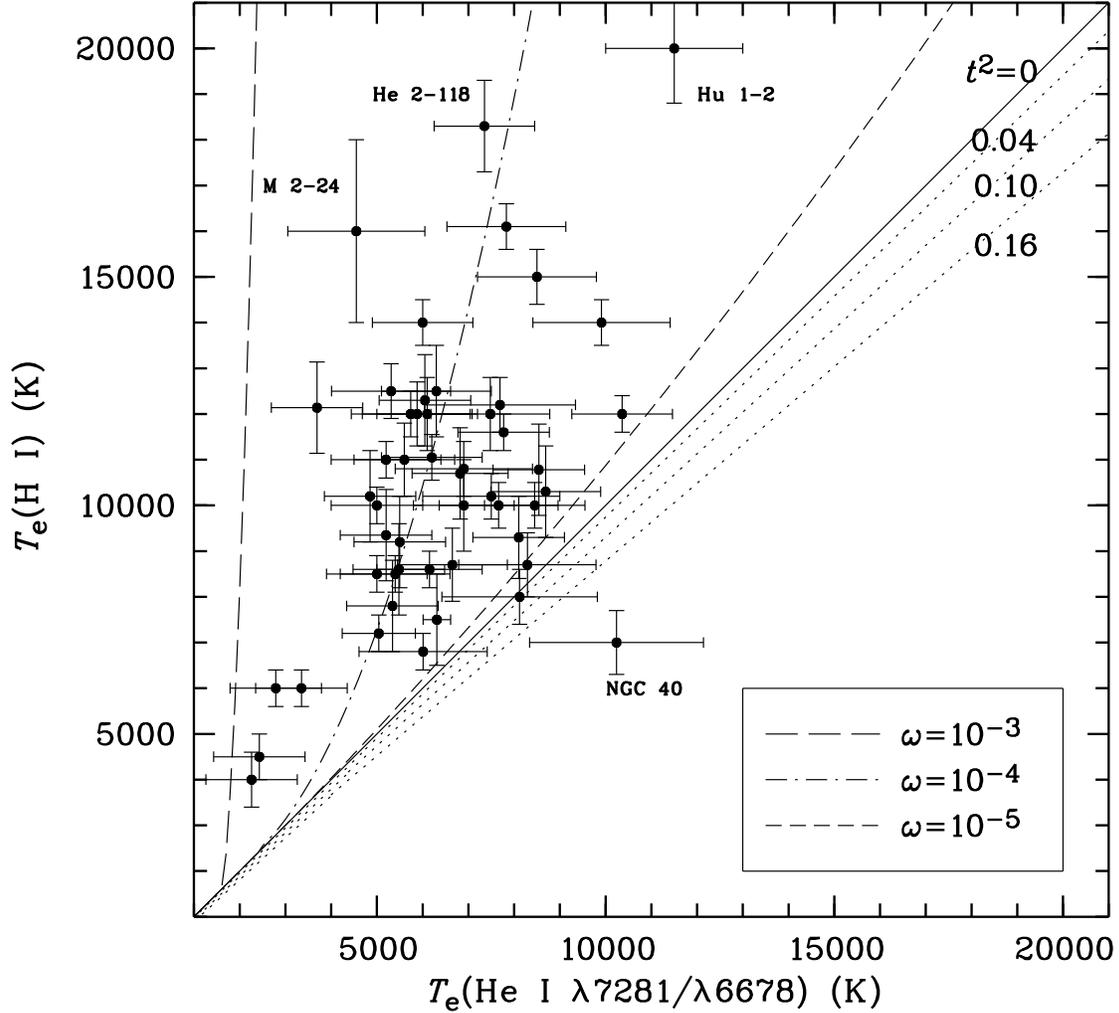,
height=14cm, bbllx=61, bblly=273, bburx=549, bbury=697, clip=, angle=0}
\caption{$T_{\rm e}$(\ion{He}{1}) versus $T_{\rm e}$(\ion{H}{1}). The solid
diagonal line is a $y=x$ plot. The dotted lines show the variations of $T_{\rm
e}$(\ion{H}{1}) as a function of $T_{\rm e}$(\ion{He}{1}) for mean square
temperature fluctuation parameter $t^2=0.04$, 0.10 and 0.16, assuming a
constant electron density of 10$^4$\,cm$^{-3}$. The short-dashed,
dot-dashed and long-dashed lines show the
variations of $T_{\rm e}$(\ion{H}{1}) as a function of $T_{\rm e}$(\ion{He}{1})
for $\omega=10^{-5}$, $10^{-4}$ and $10^{-3}$, respectively, where $\omega$ 
is the volume filling factor of H-deficient inclusions.}
\label{heihi}
\end{figure*}

The two-abundance nebular model proposed by \citet{liubarlow2000} assumes the
existence of another component of H-deficient, ultra-cold ($T_{\rm e} \sim
10^3$~K), ionized gas embedded in the diffuse gaseous nebula of ``normal''
(i.e.  $\sim$\,solar) chemical composition. In this model emission from the
H-deficient ultra-cold ionized regions has a larger contribution to the
\ion{He}{1} than to the \ion{H}{1} lines, which are still dominated by emission
from the normal component under a typical temperature of $T_{\rm e}\sim
10^4$~K.  Thus, while the model predicts that $T_{\rm e}$(\ion{H}{1}) $<$
$T_{\rm e}$([\ion{O}{3}]), it also predicts that $T_{\rm e}$(\ion{He}{1})
$<$ $T_{\rm e}$(\ion{H}{1}), in agreement with observations
(\citealt{liu209}; Fig.~\ref{heihi}).

Quantitatively, in a two-abundance nebula model, the intensity ratio of two
\ion{He}{1} lines can be written as, 
\begin{equation}
\frac{I({\rm He~{\sc I}},\lambda_1)}{I({\rm He~{\sc I}},\lambda_2)}=\frac{[N_{\rm e}N({\rm He}^+)\varepsilon_1V]_h+[N_{\rm e}N({\rm He}^+)\varepsilon_1V]_l}{[N_{\rm e}N({\rm He}^+)\varepsilon_2V]_h+[N_{\rm e}N({\rm He}^+)\varepsilon_2V]_l},
\end{equation}
where $\varepsilon_1$ and $\varepsilon_2$ are the \ion{He}{1} line emission
coefficients, $V$ is the volume of the emitting regions, and $l$ and $h$ refer
to the low-metallicity regions (i.e. the diffuse nebula of ``normal''
composition) and the high-metallicity regions (i.e. the cold H-deficient
inclusions).  Then using the analytic fit to the \ion{He}{1} line emissivity,
we have,
\begin{eqnarray}
\nonumber
\frac{I({\rm He~{\sc I}},\lambda_1)}{I({\rm He~{\sc I}},\lambda_2)}&\equiv&
\frac{a_1}{a_2}T_{\rm e4}({\rm He~{\sc I}},\lambda_1/\lambda_2)^{b_1-b_2}\exp\left[\frac{c_1-c_2}{T_{\rm e4}({\rm He~{\sc I}},\lambda_1/\lambda_2)}\right]\\
\nonumber
&=&\frac{\mu_{\rm e}\mu_{\rm He}\omega \left[a_1T_{\rm e4}^{b_1}
\exp\left(\frac{c_1}{T_{\rm e4}}\right)\right]_h+\left[a_1T_{\rm e4}^{b_1}
\exp\left(\frac{c_1}{T_{\rm e4}}\right)\right]_l}
{\mu_{\rm e}\mu_{\rm He}\omega \left[a_2T_{\rm e4}^{b_2}\exp\left(\frac{c_2}
{T_{\rm e4})}\right)\right]_h+\left[a_2T_{\rm e4}^{b_2}
\exp\left(\frac{c_2}{T_{\rm e4}}\right)\right]_l},\\
\end{eqnarray}
where $\mu_{\rm e}=(N_{\rm e})_h/(N_{\rm e})_l$ and $\mu_{\rm He}=[N({\rm
He}^+)]_h/[N({\rm He}^+)]_l$ are the electron density and He$^+$ ionic density
contrasts between the two nebular components, respectively, and
$\omega=V_h/V_l$ is the volume filling factor of the H-deficient component.
Considering that the component of ultra-cold inclusions are hydrogen-deficient,
the \ion{H}{1} recombination line and continuum emission is characterized by
the high temperature of the normal component, i.e. $T_{\rm
e}$(\ion{H}{1})$\approx (T_{\rm e})_l$.

Given $(T_{\rm e})_h$, $\mu_{\rm e}$, $\mu_{\rm He}$ and $\omega$, one can thus
obtain a relation between $T_{\rm e}$(\ion{H}{1}) and $T_{\rm e}$(\ion{He}{1}).
The current available observations are however not sufficient to provide full
constraints of all four free parameters.  Detailed photoionization models
incorporating hydrogen-deficient inclusions have been constructed by
\citet{pequignot} to account for the multi-waveband observations of PNe
\object{NGC\,6153}, \object{M\,1-42} and \object{M\,2-36}.  If we make the
simplified yet not unreasonable assumption that the postulated
hydrogen-deficient inclusions in all PNe are of similar characteristics, then
guided by the photoionization models of \citet{pequignot}, we assume that for
all PNe analyzed in the current work, $(T_{\rm e})_h=1000$~K, $\mu_{\rm e}=100$
and $\mu_{\rm He}=25$.  Under these assumptions, $T_{\rm e}$(\ion{H}{1}) as a
function of $T_{\rm e}$(\ion{He}{1}) is plotted in Fig.~\ref{heihi} for filling
factors of $\omega=0$, 10$^{-5}$, 10$^{-4}$ and 10$^{-3}$.  Here line
emissivity parameters $(a_i,b_i,c_i)$ for a density of $N_{\rm e}=10^4$ and
$10^6$\,cm$^{-3}$ have been adopted for the ``$l$'' and ``$h$'' components,
respectively.  Adopting different densities however hardly affects our results
since the emissivities of \ion{He}{1} recombination lines are almost
independent of electron density.

Fig.~\ref{heihi} shows that a small amount of hydrogen-deficient material with
a filling factor $\omega\sim10^{-4}$ can account for the observed difference
between $T_{\rm e}$(\ion{He}{1}) and $T_{\rm e}$(\ion{H}{1}) for most of the
sample PNe.  Such a small amount of hydrogen-deficient material would be
difficult to detect by direct imaging, especially in the strong background
emission of the ``normal component''. Note that in the two-abundance nebular
model, heavy element ORLs arise almost entirely from the cold H-deficient
inclusions, and their average emission temperature, $T_{\rm e}({\rm
CNONe\,ORLs})$ should represent the true electron temperature prevailing in
this H-deficient component. Thus one expects $T_{\rm e}({\rm CNONe\,ORLs})\la
T_{\rm e}$(\ion{He}{1}) \citep{liu209}, a prediction that has also been
confirmed by the available limited measurements of $T_{\rm e}({\rm
CNONe\,ORLs})$ \citep{liu209, wesson, tsamis04, liuyi04b, wesson04}. Further
accurate measurements of $T_{\rm e}({\rm CNONe\,ORLs})$ are thus essential to
better constrain the physical conditions prevailing in these regions.  For this
purpose, high S/N ratio, high spectral resolution (and preferably also high
spatial resolution) spectroscopy on a large telescope is required.

An important consideration is whether the presence of a substantial
He$^{2+}$ zone in some PNe may contribute significantly to the observed large
difference between $T_{\rm e}$(\ion{H}{1}) and $T_{\rm e}$(\ion{He}{1}).  Due
to a higher heating rate and less efficient cooling, the electron temperature
in the He$^{2+}$ zone is generally higher than that in the He$^{+}$ zone. This
is indeed the case found from photoionization models  for two PNe in our sample
(\object{NGC\,7662} and \object{NGC\,3918}) \citep[][]{harrington,clegg}. It
follows, then, that the average temperature of the H$^{+}$ zone is higher than
that of the He$^{+}$ zone since the former encompasses both the He$^{+}$ and
He$^{2+}$ zones. However, none of the photoionization models published so far
that assume a chemically homogeneous medium predict a huge $T_{\rm
e}$(\ion{H}{1}) $-$ $T_{\rm e}$(\ion{He}{1}) difference as much as 4000\,K
found in the current work.  As an example, the classic photoionization model of
\object{NGC\,7662} constructed by \citet{harrington} yields only minute values
of $t^2$(\ion{H}{1}) and $t^2$(\ion{He}{1}) and, as expected, in this model,
$T_0$(\ion{H}{1})$\approx T_{\rm e}$(\ion{H}{1}) and $T_0$(\ion{He}{1})$\approx
T_{\rm e}$(\ion{He}{1}). The model yields $T_0$(\ion{H}{1}) of 12\,590\,K, in
good agreement with the value $T_{\rm e}$(\ion{H}{1}) of $12\,200\pm600$\,K
deduced from the H~{\sc i} Balmer discontinuity.  However, their model predicts
a value of 11\,620\,K for $T_0$(\ion{He}{1}), significantly higher than the
value $7690\pm1650$\,K derived from the \ion{He}{1} $\lambda7281$/$\lambda6678$
line ratio.  Finally, we note that the observed phenomenon of large difference
between $T_{\rm e}$(\ion{H}{1}) and $T_{\rm e}$(\ion{He}{1}) persists in a
number of low excitation PNe in our sample, such as \object{H\,1-35},
\object{Hu\,2-1} and \object{M\,1-20} where the He$^{2+}$ zone is essentially
absent.  As a consequence, we conclude that the presence of a He$^{2+}$ zone in
PNe of high excitation is unlikely to play a major role in causing the large
$T_{\rm e}$(\ion{H}{1}), $T_{\rm e}$(\ion{He}{1}) discrepancy.

\begin{figure*}
\epsfig{file=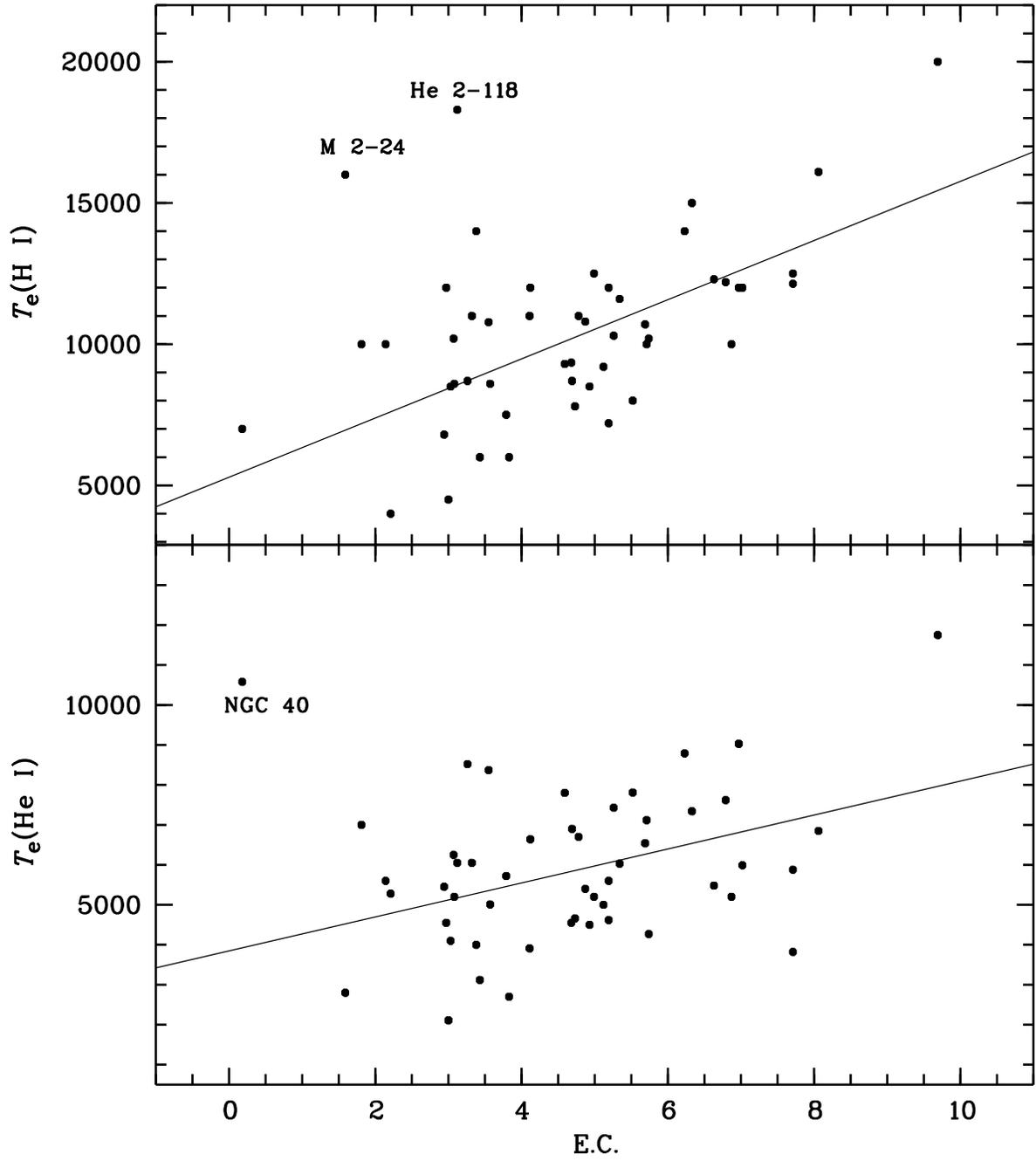,
height=18cm, bbllx=88, bblly=109, bburx=532, bbury=605, clip=, angle=0}
\caption{a) $T_{\rm e}$(\ion{H}{1}) and b) $T_{\rm e}$(\ion{He}{1})
versus Excitation Class (E.C.). The solid lines are linear least-squares
fits.}
\label{ect}
\end{figure*}

In Fig.\ref{ect}, we plot $T_{\rm e}$(\ion{H}{1}) and $T_{\rm
e}$(\ion{He}{1}) against the excitation class, E.C., calculated following the
formalism of \citet{dopita} and tabulated in Table~\ref{result}.
Fig.\,\ref{ect} shows that there is a positive correlation between $T_{\rm
e}$(\ion{H}{1}) and the E.C.. Excluding two peculiar PNe, \object{He\,2-118}
and \object{M\,2-24}, we obtain a linear correlation coefficient of 0.66. This
is consistent with the expectation that a hotter central star produces a higher
heating rate per photoionization, resulting in a higher nebular
temperature.  Fig.\ref{ect} also shows a weaker correlation between $T_{\rm
e}$(\ion{He}{1}) and the E.C..  Again after excluding \object{NGC\,40} in
addition to \object{He\,2-118} and \object{M\,2-24}, the data
yield a correlation coefficient of 0.41 between $T_{\rm e}$(\ion{He}{1}) and
the E.C.. Note that in the two-abundance nebular model, \ion{He}{1} lines are
strongly enhanced by emission from the H-deficient inclusions.  As a
consequence, the E.C. is no longer a good indicator of $T_{\rm
e}$(\ion{He}{1}). Rather, in this scenario, $T_{\rm e}$(\ion{He}{1}) will be
mainly determined by the amount and properties of the postulated H-deficient
inclusions, including their spatial distribution, density, chemical composition
and filling factor.

It is interesting to note that amongst the sample PNe the only nebula that
exhibits a higher value of $T_{\rm e}$(\ion{He}{1}) compared to $T_{\rm
e}$(\ion{H}{1}), i.e.  opposite the predictions of the two-abundance model, is
\object{NGC\,40} where $T_{\rm e}$(\ion{He}{1}) is about 3500\,K {\it higher}
than $T_{\rm e}$(\ion{H}{1}) (c.f. Fig.\ref{heihi}).  \object{NGC\,40} is
ionized by a WC8 Wolf-Rayet central star \citep{smith1969,crowther}. It is
possible that in this particular object the nebular emission line spectrum is
contaminated by emission from the strong stellar winds from the
high-temperature H-deficient envelope of the central star, leading to
apparently higher $T_{\rm e}$(\ion{He}{1}) with respect to $T_{\rm
e}$(\ion{H}{1}) \citep[c.f.][]{liuyi04b}. Further observations of this peculiar
PN are needed to clarify the situation.

\subsection{Helium abundance}

All \ion{He}{1} lines observable in the optical and UV are essentially entirely
excited by radiative recombination. As a consequence, their intensities can be
significantly enhanced by the presence of a small amount of ultra-cold,
H-deficient inclusions, leading to an overestimated overall He/H abundance
characteristic of the entire ionized region \citep{liu209,liueso}.  For a large
sample of PNe, \citet{zhang2004} showed that there is a positive correlation
between He/H abundance and the difference between the temperature derived from
the [\ion{O}{3}] forbidden lines and from the hydrogen Balmer discontinuity,
lending further support to this possibility. In the following, we present an
analytical method to obtain a quantitative estimate of the possible enhancement
of the He/H abundance due to the postulated presence of H-deficient inclusions
in PNe.  The assumptions about the two-abundance nebular model are the same as
those described above in Section~3.2. 

The intensity ratio of a \ion{He}{1} recombination line to that of H$\beta$,
$I({\rm He~{\sc I}}, \lambda_i)/I({\rm H}\beta)$ observed from a nebula is
given by
\begin{eqnarray}
\nonumber
\frac{I({\rm He~{\sc I}}, \lambda_i)}{I({\rm H}\beta)}& = &\frac{N({\rm He}^+, \lambda_i)(\varepsilon_i)_l4861}{N({\rm H}^+)(\varepsilon_{{\rm H}\beta})_l\lambda_i}\\
&=&\frac{[N_{\rm e}N({\rm He}^+)\varepsilon_iV]_h+[N_{\rm e}N({\rm He}^+)\varepsilon_iV]_l}{[N_{\rm e}N({\rm H}^+)\varepsilon_{{\rm H}\beta}V]_h+[N_{\rm e}N({\rm H}^+)\varepsilon_{{\rm H}\beta}V]_l}\frac{4861}{\lambda_i},
\end{eqnarray}
where $N({\rm He}^+, \lambda_i)/N({\rm H}^+)$ is the He$^+$/H$^+$ ionic
abundance ratio determined assuming a chemically homogeneous nebula.  From the 
above equation we find,
\begin{equation}
\frac{N({\rm He}^+, \lambda_i)}{N({\rm H}^+)}\approx\left[\omega\mu_{\rm e}\mu_{\rm He}\frac{(\varepsilon_i)_h}{(\varepsilon_i)_l}+1\right]\left[\frac{N({\rm He}^+)}{N({\rm H}^+)}\right]_l.
\end{equation}

For the small amount of H-deficient material as hypothesized in the
two-abundance model, the average He$^+$/H$^+$ ionic abundance ratio for the
entire nebula is essentially identical to the value for the diffuse material,
i.e.  $[N({\rm He}^+)/N({\rm H}^+)]_l$. In such a case, the traditional method
of abundance analysis assuming a chemically homogeneous nebula will
overestimate the He/H abundance by a factor of [$\omega\mu_{\rm e}\mu_{\rm
He}(\varepsilon_i)_h/(\varepsilon_i)_l+1$].  For the representative values of
$\mu_{\rm e}=100$, $\mu_{\rm He}=25$ and $\omega=10^{-4}$, the He$^+$/H$^+$
ionic abundance ratio will be overestimated by a factor of 1.25.  For example,
the empirical analysis of NGC\,6153 by \citet{liubarlow2000} assuming a
chemically homogeneous nebula yields a He/H elemental abundance ratio of 0.136,
which is a factor of 1.35 higher than the average He/H abundance ratio for the
entire ionized region derived from the detailed two-abundance photoionization
modeling of this nebula by \citet{pequignot}. The much lower He/H abundance
ratio obtained by two-abundance photoionization modeling is supported by
observations of the collisional excitation dominated \ion{He}{1} 2s\,$^3$S --
2p\,$^3$P$^{\rm o}$ $\lambda$10830 line \citep{liu209}.

Many PNe are known to show enhancement of helium with respect to the Sun. For
example, the average He/H ratio obtained by \citet{kingsburgh} for a large
sample of Galactic PNe is a factor of 1.35 solar. The enhancement is often
ascribed to the second and third dredge-up processes that occur during the
post-main-sequence evolution stages of the progenitor stars of PNe. However,
the extremely high He abundances deduced for PNe such as \object{He~2-111}
\citep{kingsburgh} are difficult to explain by the current theories of
nucleosynthesis and dredge-up processes. It is possible therefore that the very
high He/H abundances observed in those extreme He-rich PNe are actually caused
by the contribution of H-deficient inclusions embedded in the nebulae.

The CEL/ORL abundance discrepancy and the BJ/CEL temperature discrepancy which
are ubiquitously observed amongst PNe are also found in \ion{H}{2} regions
\citep[e.g.][]{tsamisb}. The determination of $T_{\rm e}$(\ion{H}{1}) for
\ion{H}{2} regions is much more difficult than for PNe due to the generally
much lower surface brightness of H~{\sc ii} regions and the strong
contamination of scattered star light to the nebular spectrum in dusty H~{\sc
ii} regions.  Recently, \citet{garc} presented deep echelle spectroscopy of the
Galactic \ion{H}{2} region \object{NGC\,3576}, covering the wavelength range
from 3100--10400\,{\AA}.  They derived $T_{\rm e}$(\ion{H}{1}) of
$6650\pm750$\,K and $T_{\rm e}$([\ion{O}{3}]) of $8500\pm50$~K, in agreement
with the general pattern observed amongst PNe that $T_{\rm e}$(\ion{H}{1}) is
systematically lower than $T_{\rm e}$([\ion{O}{3}]). On the other hand, from
the \ion{He}{1} $\lambda\lambda6678,7281$ line fluxes reported in their paper,
we obtain $T_{\rm e}$(\ion{He}{1}) of $6800\pm600$\,K, which is consistent with
$T_{\rm e}$(\ion{H}{1}) within the errors. Note that \object{NGC\,3576} has a
relatively small ORL/CEL abundance discrepancy (a factor of 1.8;
\citealt{tsamis,garc}). Measurements of $T_{\rm e}$(\ion{H}{1}) and $T_{\rm
e}$(\ion{He}{1}) in more H~{\sc ii} regions are needed, especially for those
exhibiting large ORL/CEL abundance discrepancies in order to have a better
picture of this problem in H~{\sc ii} regions.

For a few metal poor extragalactic \ion{H}{2} regions, \citet{peimbert2002}
find that the electron temperatures derived from \ion{He}{1} recombination
lines are also systematically lower than the corresponding values deduced from
the [\ion{O}{3}] forbidden line ratio.  If we ascribe the discrepancy to the
presence of H-deficient inclusions in these nebulae, then the effects of these
inclusions on the determination of the primordial helium abundance can be
estimated roughly from Eqs.~(13) and (15). Using the line fluxes presented by
\citet{peimbert2002} and assuming that the H-deficient material in these
metal-poor H~{\sc ii} regions have similar composition as those postulated to
exist in PN \object{NGC\,6153}, we estimate that the primordial He abundance
could have been overestimated by as much as 5\%.

\section{Conclusions}

In this paper, we show that \ion{He}{1} recombination lines can be used to
probe nebular thermal structures and provide vital information regarding the
nature and physical causes of the long-standing ORL/CEL temperature and
abundance determination discrepancies. We present electron temperatures derived
from the \ion{He}{1} $\lambda7281/\lambda6678$ ratio for 48 PNe. The
\ion{He}{1} temperatures are found to be systematically lower than those
deduced from the Balmer jump of the \ion{H}{1} recombination spectrum. The
result is exactly opposite to the predictions of the scenarios of temperature
fluctuation and density inhomogeneities but in good agreement with the
expectations of the two-abundance nebular model proposed by
\citet{liubarlow2000}.  We estimate that a filling factor of the order of
$10^{-4}$ of the ultra-cold, H-deficient material is sufficient to explain the
observed differences between the \ion{H}{1} and \ion{He}{1} temperatures.  We
show that the possible presence of H-deficient inclusions in PNe may indicate
that current estimates of He/H abundances in PNe may have been overestimated by
a factor of $\sim1.25$.  The possible existence of hydrogen-deficient
inclusions in \ion{H}{2} regions could also cause the primordial helium
abundance determined from metal-poor \ion{H}{2} galaxies to be overestimated.
We stress the importance of high S/N ratio, high resolution spectroscopy in 
the tackling of these fundamental problems.

\acknowledgments

We thank the referee, Dr. V. Luridiana, for helpful suggestions and incisive
comments that improved the paper significantly.  YZ acknowledges the award of a
Liu Yongling Scholarship.  Support for RHR was from the NASA Long-Term Space
Astrophysics (LTSA) program.

\appendix

\section{\ion{He}{1} line emissivities ($T_{\rm e}<5000$\,K)}

Combining the \ion{He}{1} recombination model of  \citet{smits1996} and the
collisional excitation rates of \citet{sawey}, we have calculated emissivities
for the \ion{He}{1} $\lambda\lambda$4471, 5876, 6678 and 7281 lines for $T_{\rm
e}<5000$\,K -- the temperature range not considered by \citet{benjamin}.  At
such low temperatures, the effects of collisional excitation are however
essentially negligible. We may write, following the analytic expression used by
\citet{benjamin}, 
\begin{equation}
\varepsilon_i=a_iT_{\rm e4}^{b_i}\exp(c_i/T_{\rm e4})\,{\rm erg}\,{\rm cm}^3\,{\rm s}^{-1},
\end{equation}
where $T_{\rm e4}=T_{\rm e}/10^4$\,K, and $a_i$, $b_i$ and $c_i$ are constants.  The
constants were derived using a least-squares algorithm. The results are
tabulated in Table~\ref{fit} for electron densities of $N_{\rm e}=10^2$, $10^4$ and
$10^6$~cm$^{-3}$.  The maximum errors of the fits for the temperature
range $312.5-5000$\,K, $e_i$, are also listed in 
the last column of the Table.

\clearpage

\begin{deluxetable}{lccccc}
\tablecaption{Electron temperatures derived
from \ion{He}{1} line ratios.\label{result}}
\tablewidth{0pt}
\tablehead{
\colhead{Source} & 
\colhead{$T_{\rm e}$(He~{\sc i} $\lambda7281/\lambda6678$)} &
\colhead{$T_{\rm e}$(He~{\sc i} $\lambda7281/\lambda5876$)} &
\colhead{$T_{\rm e}$(\ion{H}{1})} & E.C. &\colhead{Ref.} 
}
\startdata
Cn 2-1       & $ 6900\pm1500$ & $ 5400\pm1300$ &  $10800\pm 600$ &4.87&...\\
H 1-35       & $ 6900\pm1100$ & $ 5600\pm1200$ &  $10000\pm1000$ &2.14&...\\
H 1-50       & $ 6300\pm1200$ & $ 5200\pm1300$ &  $12500\pm1000$ &4.99&...\\
He 2-118     & $ 7350\pm1100$ & $ 6050\pm1100$ &  $18300\pm1000$ &3.12&...\\
Hu 1-2       & $11500\pm1500$ & $11750\pm3000$ &  $20000\pm1200$ &9.69& L04\\
Hu 2-1       & $ 8400\pm1100$ & $ 7000\pm1200$ &  $10000\pm 500$ &1.81& W04\\
IC 1297      & $ 5000\pm1000$ & $ 5200\pm1000$ &  $10000\pm 400$ &6.87&...\\
IC 2003      & $ 5600\pm1100$ & $ 6700\pm1200$ &  $11000\pm 800$ &4.78& W04\\
IC 3568      & $ 8100\pm1000$ & $ 7800\pm1450$ &  $ 9300\pm 900$ &4.59& L04\\
IC 4191      & $ 5500\pm1000$ & $ 5000\pm1000$ &  $ 9200\pm1000$ &5.12& T03\\
IC 4406      & $ 5200\pm1000$ & $ 4550\pm1000$ &  $ 9350\pm1000$ &4.68& T03\\
IC 4634      & $ 5400\pm1200$ & $ 4100\pm1000$ &  $ 8500\pm 400$ &3.03&...\\
IC 4776      & $ 6150\pm1150$ & $ 5200\pm1300$ &  $ 8600\pm 400$ &3.08&...\\
IC 4997      & $ 7500\pm1500$ & $ 6250\pm1500$ &  $10200\pm 500$ &3.07&...\\
IC 5217      & $ 6100\pm1100$ & $ 5600\pm1200$ &  $12000\pm 800$ &5.19& W04\\
M 1-20       & $ 5880\pm1200$ & $ 4550\pm1000$ &  $12000\pm 700$ &2.97&...\\
M 1-42       & $ 2260\pm1000$ & $ 5280\pm1200$ &  $ 4000\pm 600$ &2.21&...\\
M 2-24       & $ 4550\pm1500$ & $ 2800\pm1000$ &  $16000\pm2000$ &1.59&...\\
M 2-36       & $ 2790\pm1000$ & $ 3120\pm1000$ &  $ 6000\pm 400$ &3.43&...\\
M 3-21       & $ 5200\pm1200$ & $ 3910\pm1000$ &  $11000\pm 400$ &4.11&...\\
M 3-32       & $ 2430\pm1000$ & $ 2110\pm1000$ &  $ 4500\pm 500$ &3.00&...\\
Me 2-2       & $ 6200\pm1100$ & $ 6050\pm1200$ &  $11000\pm 500$ &3.32& W04\\
NGC 40       & $10240\pm1900$ & $10580\pm4200$ &  $ 7000\pm 700$ &0.18& L04\\
NGC 3132     & $ 8540\pm1000$ & $ 8370\pm1000$ &  $10780\pm1000$ &3.55& T03\\
NGC 3242     & $ 4850\pm1000$ & $ 4270\pm1000$ &  $10200\pm1000$ &5.74& T03\\
NGC 3918     & $ 6050\pm1000$ & $ 5480\pm1000$ &  $12300\pm1000$ &6.63& T03\\
NGC 5307     & $ 6820\pm1050$ & $ 6540\pm1100$ &  $10700\pm1000$ &5.69& R03\\
NGC 5315     & $ 5480\pm1000$ & $ 5010\pm1000$ &  $ 8600\pm1000$ &3.57& T03\\
NGC 5315     & $ 6310\pm 300$ & $ 5720\pm 400$ &  $ 7500\pm1000$ &3.79& P04\\
NGC 5873     & $ 5740\pm1300$ & $ 5990\pm1400$ &  $12000\pm 500$ &7.02&...\\
NGC 5882     & $ 5340\pm1000$ & $ 4660\pm1000$ &  $ 7800\pm1000$ &4.73& T03\\
NGC 6153     & $ 3350\pm1000$ & $ 2700\pm1000$ &  $ 6000\pm 400$ &3.83&...\\
NGC 6210     & $ 6650\pm1200$ & $ 6900\pm1350$ &  $ 8700\pm 800$ &4.69& L04\\
NGC 6302     & $ 7830\pm1300$ & $ 6850\pm1300$ &  $16100\pm 500$ &8.06&...\\
NGC 6543     & $ 6010\pm1400$ & $ 5450\pm1400$ &  $ 6800\pm 400$ &2.94&...\\
NGC 6567     & $ 7480\pm1300$ & $ 6640\pm1300$ &  $12000\pm 800$ &4.12&...\\
NGC 6572     & $ 8690\pm1200$ & $ 7430\pm1650$ &  $10300\pm1000$ &5.26& L04\\
NGC 6620     & $ 7660\pm1300$ & $ 7120\pm1300$ &  $10000\pm 500$ &5.71&...\\
NGC 6720     & $ 8120\pm1700$ & $ 7810\pm2200$ &  $ 8000\pm 600$ &5.52& L04\\
NGC 6741     & $ 8500\pm1300$ & $ 7340\pm1880$ &  $15000\pm 600$ &6.33& L04\\
NGC 6790     & $ 9910\pm1500$ & $ 8790\pm2000$ &  $14000\pm 500$ &6.23& L04\\
NGC 6803     & $ 5000\pm1100$ & $ 4500\pm1200$ &  $8500 \pm 400$ &4.93& W04\\
NGC 6833     & $ 6000\pm1100$ & $ 4000\pm1200$ &  $14000\pm 500$ &3.38& W04\\
NGC 6818     & $ 3690\pm1000$ & $ 3820\pm1000$ &  $12140\pm1000$ &7.71& T03\\
NGC 6818     & $ 5310\pm1300$ & $ 5880\pm1300$ &  $12500\pm 600$ &7.71&...\\
NGC 6826     & $ 8290\pm1500$ & $ 8520\pm2000$ &  $ 8700\pm 700$ &3.26& L04\\
NGC 6884     & $ 7770\pm1000$ & $ 6030\pm1550$ &  $11600\pm 400$ &5.34& L04\\
NGC 7009     & $ 5040\pm 800$ & $ 4620\pm1100$ &  $ 7200\pm 400$ &5.19&...\\
NGC 7027     & $10360\pm1100$ & $ 9030\pm2200$ &  $12000\pm 400$ &6.97&...\\
NGC 7662     & $ 7690\pm1650$ & $ 7620\pm1950$ &  $12200\pm 600$ &6.79& L04\\
\enddata
\tablerefs{ [L04] \citet{liuyi04a}; [P04] \citet{peimbert2004};
[R03] \citet{ruiz2003}; [T03] \citet{tsamis}; [W04] \citet{wesson04}
}
\end{deluxetable}

\begin{deluxetable}{ccccc}
\tablecaption{Fitting parameters for \ion{He}{1} line 
emissivities.\label{fit}}
\tablewidth{0pt}
\tablehead{
\colhead{Emissivities} & \colhead{$a_i$}
 &
\colhead{$b_i$} &
\colhead{$c_i$} &
\colhead{$e_i$}
}
\startdata
\multicolumn{5}{l}{$N_{\rm e}$=10$^2$~cm$^{-3}$} \\
$j_{4471}$ & 6.835$\times10^{-26}$& -0.8224& -0.0074& $1.40\%$ \\
$j_{5876}$ & 1.838$\times10^{-25}$& -0.9745& -0.0086& $1.11\%$ \\
$j_{6678}$ & 5.251$\times10^{-26}$& -0.9819& -0.0088& $1.24\%$ \\
$j_{7281}$ & 9.104$\times10^{-27}$& -0.5594&  0.0007& $0.25\%$ \\
\multicolumn{5}{l}{$N_{\rm e}$=10$^4$~cm$^{-3}$}\\
$j_{4471}$ & 6.775$\times10^{-26}$& -0.8270& -0.0041& $1.50\%$ \\
$j_{5876}$ & 1.824$\times10^{-25}$& -0.9702& -0.0068& $1.14\%$ \\
$j_{6678}$ & 5.203$\times10^{-26}$& -0.9767& -0.0068& $1.23\%$ \\
$j_{7281}$ & 9.463$\times10^{-27}$& -0.5289&  0.0068& $1.14\%$ \\
\multicolumn{5}{l}{$N_{\rm e}$=10$^6$~cm$^{-3}$}\\
$j_{4471}$ & 6.746$\times10^{-26}$& -0.8337&  0.0067& $1.23\%$ \\
$j_{5876}$ & 1.819$\times10^{-25}$& -0.9509&  0.0035& $1.37\%$ \\
$j_{6678}$ & 5.185$\times10^{-26}$& -0.9581&  0.0035& $1.67\%$ \\
$j_{7281}$ & 9.663$\times10^{-27}$& -0.5307&  0.0179& $2.29\%$ \\
\enddata
\end{deluxetable}

\end{document}